\documentclass[twocolumn,superscriptaddress,prl,longbibliography]{revtex4-1}
\usepackage{graphicx}
\usepackage{amssymb,amsmath,amsfonts,bm}
\usepackage{hyperref}
\usepackage[dvipsnames]{xcolor}
\hypersetup{urlcolor=RoyalBlue, colorlinks=true,citecolor=RoyalBlue,linkcolor=RoyalBlue} 
\usepackage{physics}
\usepackage[final]{showkeys}
\usepackage{float}
\begin{document}
\title{Melting of three-sublattice order in triangular lattice Ising antiferromagnets: Power-law order, $Z_6$ parafermionic multicriticality, and weakly first order transitions}
\author{Geet Rakala}
\affiliation{\small{Okinawa Institute of Science and Technology Graduate University, Onna-son, Okinawa-ken 904-0412, Japan}}
\author{Nisheeta Desai}
\affiliation{\small{Tata Institute of Fundamental Research, 1 Homi Bhabha Road, Mumbai 400005, India}}
\author{Saumya Shivam}
\affiliation{\small{Physics Department, Princeton University, Princeton, NJ 08544, U.S.A}} 
\author{Kedar Damle}
\affiliation{\small{Tata Institute of Fundamental Research, 1 Homi Bhabha Road, Mumbai 400005, India}}

\begin{abstract}
The nature of the thermal melting process by which triangular-lattice Ising antiferromagnets lose their low-temperature ferrimagnetic three-sublattice order depends on the range of the interactions: It changes character when second and third neighbour ferromagnetic interactions become comparable to the nearest-neighbour antiferromagnetic coupling. We present a detailed numerical characterization of the corresponding threshold at which two-step melting of three-sublattice order gives way to a direct first-order transition at which this order is lost. The multicritical behaviour at this threshold is argued to be in the universality class of the $Z_6$ parafermion conformal field theory with central charge $c=5/4$. The presence of this multicritical threshold influences the melting behaviour and long-wavelength properties over a fairly large range of parameters, and at temperatures that are of the same order as the exchange interactions. It is therefore of potential experimental relevance in the context of easy-axis triangular lattice antiferromagnets that display such low temperature ordering.

\end{abstract}

\maketitle
{\it Introduction:} Low temperature phases of condensed matter systems are often characterized by spontaneous symmetry breaking and long-range order. The nature of the melting transition at which this order is lost on heating is usually understood in terms of the Landau-Ginzburg theory for such phase transitions. In this framework, the nature of the melting transition is predicted by the form of the Landau free energy functional, written in terms of the order parameter field; this form is, in turn, largely determined by the demands of global symmetry, analyticity, and locality~\cite{goldenfeld_book}. If a continuous transition is a generic possibility within this framework, such an analysis usually pins down the universality class of the transition and serves as the basis for more sophisticated Wilsonian renormalization group predictions of scaling behaviour associated with such critical points.

In some unusual examples~\cite{chern2012MagneticCharge,damle2015MeltingThreeSublattice}, this framework admits more than one generic possibility for continuous melting transitions. In such a situation, the nature of the melting process can change in interesting ways as a function of system parameters, although the character of the low temperature state remains qualitatively unchanged. In particular, such systems can display multicritical behaviour associated with such changes in the character of the melting transition. Here, we present detailed computational evidence for such multicriticality in the phase diagram of frustrated Ising antiferromagnets with ferromagnetic further neighbour couplings that extend up to third neighbours on the triangular lattice. Our study of such multicritical behaviour is interesting from two quite distinct points of view, which we discuss next.

The first has to do with the storied success of field-theoretical ideas in understanding two-dimensional critical phenomena on the basis of emergent conformal invariance at criticality~\cite{senechal_book,friedan1984ConformalInvariance}. While this approach leads to an essentially complete classification of all two-dimensional critical phenomena described by conformal field theories (CFTs) with central charge $c<1$~\cite{senechal_book}, a detailed understanding of  $c > 1$ CFTs and the corresponding critical phenomena is largely limited to cases in which the CFT possesses an enlarged symmetry group that can be leveraged to obtain a complete characterization of the critical behaviour. Examples of this include the discrete series of self-dual $Z_N$-symmetric CFTs~\cite{zamolodchikov1985NonlocalParafermion} with $N=4,5\dots$, which possess an an enlarged `parafermion' symmetry algebra~\cite{Fendley_freeparafermions,Fendley_parafermion_latticeoperators}, and the supersymmetric CFTs~\cite{friedan1985SuperconformalInvariance} $SM_p$ with $p = 4,5\dots$. Viewed from this vantage point, the key message from our work is that the multicritical behaviour of triangular lattice antiferromagnets studied here provides a microscopic realization of the $Z_6$ self-dual parafermionic CFT which also admits an alternate description as the supersymmetric CFT $SM_6$ and has central charge $c=5/4$.

\begin{figure*}
    \centering
    \includegraphics[width=\textwidth]{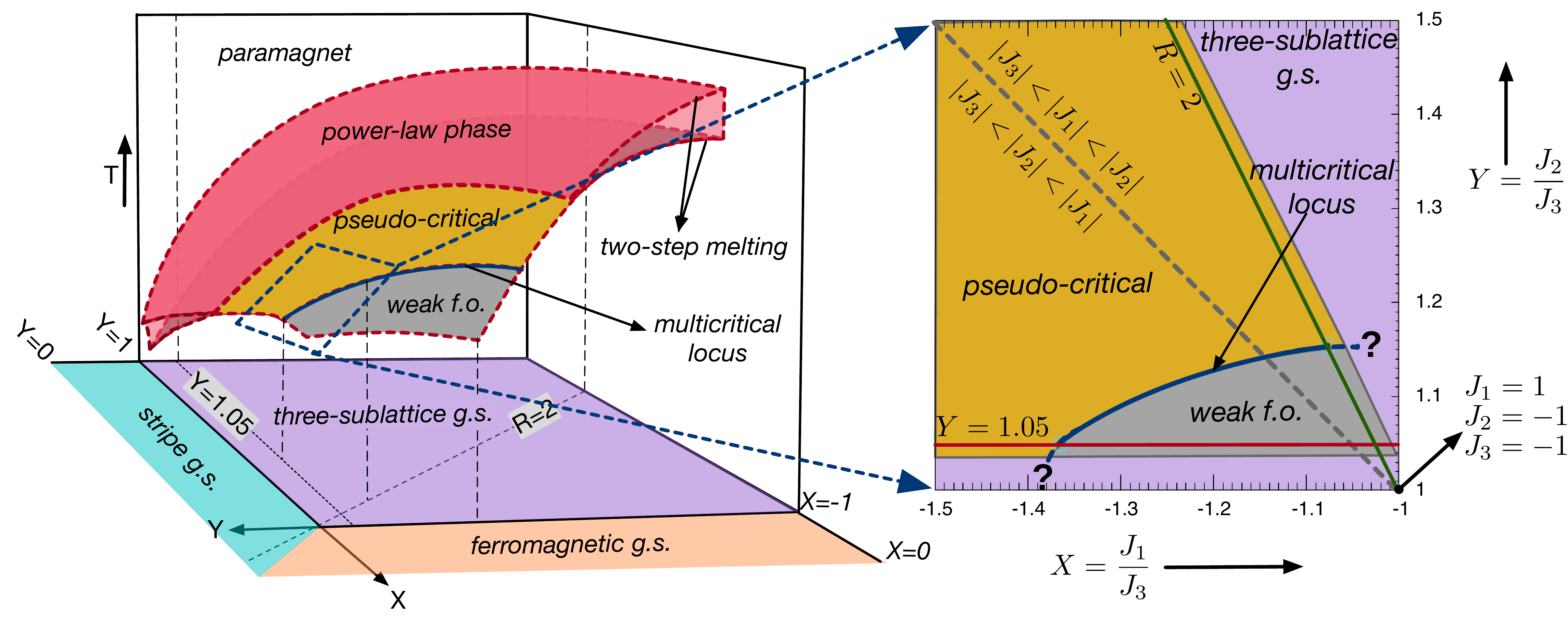}
    \caption{ The figure on the left shows a schematic of the 3D melting phase diagram in the $X=\frac{J_1}{J_3},Y=\frac{J_2}{J_3}$ and $T$ space for $J_3<0$ (ferromagnetic). The three-sublattice ground state is stabilised when $X<-1$ and $Y>1$. We set $J_1=+1$ (antiferromagnetic) to fix the overall energy scale and study the melting phase diagram along the $R \equiv -(J_2+J_3)/J_1 = 2$ and $Y=1.05$ lines. The figure on the right shows a true to scale projection of the melting phase diagram above the $XY$ plane. In the right panel, the dashed line satisfies $Y=-X$, so $|J_3|<|J_2|<|J_1|$ ($|J_3|<|J_1|<|J_2|$) below (above) this line. We find that the finite-temperature critical phase associated with two-step melting pinches off along a multicritical locus if one approaches the point $(X,Y) = (-1, 1)$. Upon crossing this multicritical locus, the melting transition becomes weakly first order. The coordinates of the multicritical point on the $Y=1.05$ line ($R=2$ line) are given approximately by $X=-1.365$, $Y=1.05$ ($X=-1.0767$, $Y=1.1534$).}
    \label{fig:phase_diagram}
\end{figure*}

Second, as we now discuss, this multicriticality is of potential experimental relevance in the context of triangular lattice antiferromagnets with strong easy-axis anisotropy. The low temperature physics of such Ising antiferromagnets is expected to be characterized by long-range ferrimagnetic ordering~\cite{landau1983CriticalMulticritical} at the three-sublattice wavevector ${\mathbf Q} = (2\pi/3,2\pi/3)$ of the triangular lattice for a wide range of relative strengths of the first, second and third neighbour couplings. This ordered state spontaneously develops a uniform magnetization as well as a spin modulation at wavevector ${\mathbf Q}$. A Landau-Ginzburg analysis suggests that such three-sublattice order can melt on heating in multiple ways~\cite{damle2015MeltingThreeSublattice}. Apart from a direct first order transition, which is of course possible, three-sublattice order can also be lost via a two-step melting process with an intermediate phase with power-law three-sublattice order, or via a sequence of two second-order transitions, the first being a transition in the three-state Potts universality class associated with the loss of the spin modulation at wavevector ${\mathbf Q}$, and a subsequent higher temperature transition in the Ising universality class associated with the loss of the uniform magnetization.

The multicritical behaviour studied here corresponds to the threshold at which two-step melting of three-sublattice order gives way to a single first-order melting transition. Accessing this multicritical point naturally requires a degree of fine-tuning: Three-sublattice order is lost via such multicritical melting only when the values of the nearest-neighbour antiferromagnetic interaction ($J_1 > 0$) and the second and third neighbour ferromagnetic interactions ($J_2 <0$ and $J_3 < 0$ respectively) place the system along a one-dimensional locus in the second quadrant of the $J_1/J_3$, $J_2/J_3$ plane. 
However, and this bring us to the second interesting aspect of our results, this multicriticality influences the long-wavelength behaviour of various correlation functions is over a much wider range of parameters. 

Moreover, the multicritical locus itself can be accessed without needing unphysically large values of further-neighbour interactions: Indeed, part of this locus is in the regime with roughly comparable second and third-neighbour couplings, both of which are slightly smaller than the dominant nearest-neighbour antiferromagnetic interaction, making experimental realizations that much more plausible. Taken together, these two features throw open the possibility that this interesting multicritical phenomenon, which is characterized by emergent supersymmetry, may also be accessible to experiments on appropriately chosen insulating antiferromagnets with strong easy-axis anisotropy. This motivates our detailed computational study of the melting of three-sublattice order in such triangular lattice Ising antiferromagnets.

{\em Overview of results and conclusions:}
We consider the classical triangular lattice Ising antiferromagnet with nearest neighbour antiferromagnetic exchange $J_1>0$ and ferromagnetic second and third neighbour exchange couplings $J_2 < 0$ and $J_3 < 0$, {\em i.e.} with classical Hamiltonian:
\begin{eqnarray}
\label{eq:hamiltonian}
H&=&J_1\sum_{\langle rr' \rangle} \sigma_r \sigma_{r'} + J_2 \sum_{\langle \langle rr' \rangle \rangle} \sigma_r \sigma_{r'} + J_3  \sum_{\langle \langle \langle rr' \rangle \rangle \rangle} \sigma_r \sigma_{r'} \nonumber \\
&&
\end{eqnarray}
where $\langle r r' \rangle$, $\langle \langle r r' \rangle \rangle$, and $\langle \langle \langle r r' \rangle \rangle \rangle$ denote nearest neighbor, next-nearest neighbor,
and next-next-nearest neighbor links of the triangular lattice, and $\sigma_r = \pm 1$
are the Ising spins at sites $r$ of this lattice. When $J_2=J_3=0$, there is no long-range order at any temperature, and the ground state ensemble has power-law three-sublattice order. Ferromagnetic $J_2 < 0$ stabilizes long-range ferrimagnetic three-sublattice order, while ferromagnetic $J_3$ stabilizes a striped order in the ground state. For $J_3 < 0$, the complete ground state phase diagram as a function of $X=J_1/J_3$ and $Y=J_2/J_3$ has been summarized earlier~\cite{tanaka1975GroundState}. 

Here, we confine ourselves to a part of the second quadrant  given by the inequalities $X<-1$ and $Y>1$, for which the ground state has three-sublattice order (see Fig.~\ref{fig:phase_diagram}). 
To fix the overall energy scale, we set $J_1=1$ and measure the temperature and all other couplings in these units. We study the thermal melting of three-sublattice order for various cuts in this region of the  $(X Y)$ plane. Our most detailed data sets are along the line $R \equiv -(J_2+J_3)/J_1 = 2$ parameterized by the coordinate $\kappa = -(J_2-J_3)/J_1$, and the line $Y=1.05$, parameterized by the coordinate $X$. 

On the $R=2$ line, we find that the three sublattice order melts via a two-step melting process for $\kappa \geq 0.2$. For $0.2 > \kappa \gtrsim 0.1425$, we find that the three-sublattice order melts via a pseudo-critical transition which we discuss in detail below. This pseudo-critical region along the $R=2$ line ends at a multicritical point at $\kappa \approx 0.1425$. For $\kappa \lesssim 0.1425$, we find that the melting transition is weakly first order. On the $Y=1.05$ line, we find a clear two-step melting behaviour at $X=-8$, pseudocritical behavior slightly to the left of a multicritical end-point at $X \approx-1.3650$, and a weak first order melting transition for $X \gtrsim -1.3650$. These results, and the results of other less detailed scans, allow us to conclude that there is a locus of multicritical points at which the finite-temperature power-law ordered critical phase associated with two-step melting pinches off, with weakly first order melting behaviour on the other side of this locus (Fig.\ref{fig:phase_diagram}). Our central result is that this multicritical locus is in the universality class of the self-dual $Z_6$ parafermion CFT mentioned earlier.
In the remainder of this paper, we provide the evidence and analysis that leads us to these conclusions.

{\em Observables: }
To identify and characterize the two-step melting region, we define a local complex three-sublattice order parameter $\psi(r) \equiv |\psi|(r)e^{i \theta(r)}$ along the lines of Ref.~\onlinecite{damle2015MeltingThreeSublattice} as
\begin{eqnarray}
\psi(r) &=& \sigma_a(r) +\sigma_b(r)\exp(2\pi i/3) + \sigma_c(r)  \exp(4 \pi i /3) \nonumber \\
&&
\end{eqnarray}
where the subscripts refer to the three-sublattice decomposition of the triangular lattice into $a$, $b$ and $c$ sublattices, and the three spins being summed over all belong to an up-pointing triangle whose $a$ sublattice site is at $r$. Note that $\psi^3(r) \equiv \sigma_a(r) + \sigma_b(r) +\sigma_c(r)$ is the local magnetization density which we denote with a slight abuse of notation by $\sigma(r)$, while $\psi^2(r)$ represents an independent linear combination of the three spins belonging to the up-pointing triangle at $r$. We measure the two point correlation functions
\begin{equation}
C_{\mathcal{O}}(\vec{r}) = \langle {\mathcal O}^{*}(r) \cdot {\mathcal O}(0) + {\rm h.c.} \rangle 
\end{equation} 
for ${\mathcal O}(r)$ equal to $\psi(r)$, $\psi^2(r) \equiv \phi(r)$ and $\psi^3(r) \equiv \sigma(r)$. 
Viewing $\psi$ as a two-component order parameter, and $\sigma$ as a one-component scalar, we also measure the corresponding Binder cumulants, defined for an $n$-component order-parameter as  
$U_{\mathcal{O}}(L) = 1- \frac{n}{(n+2)}\frac{\langle |\mathcal{O}|^4 \rangle}{\langle |\mathcal{O}|^2 \rangle^2}$. As $L \rightarrow \infty$, $U_{\mathcal O}(L)$ tends to $0$ ($\frac{2}{n+2}$) in a phase in which ${\mathcal O}$ is disordered (ordered). At criticality $U_{\mathcal O}(L)$ is expected to take on a universal $L$-independent value. 


{\em Perspective from Landau theory:} The simplest Landau theory treatment\cite{landau1983CriticalMulticritical,damle2015MeltingThreeSublattice} of three-sublattice ordering is in terms of the order parameter $\psi(r)$. Since the critical behavior is expected to be controlled by the phase fluctuations of $\psi$, this reduces to a Landau theory written in terms of the local phase $\theta(r)$. The Landau theory action for $\theta(r)$ has six-fold anisotropy which favours the six values $2\pi m/6$ ($m=0,1\dots 5$). If we use a lattice discretization of this theory and incorporate the six-fold anisotropy by working with a clock variable $\theta$ that only takes on the six values $2\pi m/6$, this Landau theory reduces to the generalized six-state clock model with classical Hamiltonian $H = \sum_{\langle r r' \rangle} V(\theta_r - \theta_r')$
where the sum is over nearest neighbor pairs of the lattice~\cite{cardy1980GeneralDiscrete}.
The potential $V$ can be written as:
\begin{equation}
V(x) = K_1 [1-\cos(x)] + K_2 [1-\cos(2x)] + K_3 [1-\cos(3x)].
\end{equation}
From the Landau theory point of view, $\theta_r$ can be identified with the phase of the complex three-sublattice order parameter $\psi(r)$ defined earlier.

At low temperatures, the six-state clock model is ferromagnetically ordered, with $\theta$ freezing into one of its six possible values. This six-fold symmetry breaking can melt either via a two-step melting transition with an intermediate power-law ordered phase~\cite{jose1977RenormalizationVortices} , or via a first order transition~\cite{cardy1980GeneralDiscrete}, or via a sequence of two second order transitions. 
In the intermediate power-law ordered region associated with two-step melting, correlators of $\exp(i\theta)$, 
$\exp(2 i \theta)$ and $\exp(3i\theta)$ all show power law behavior, with anomalous exponent $\eta_{\theta} \in [1/9,1/4]$ for $T \in [T_{c1},T_{c2}]$, $\eta_{2 \theta} = 4\eta_{\theta}$  and $\eta_{3\theta} = 9 \eta_{\theta}$~\cite{jose1977RenormalizationVortices}. In the generalized six-state clock model, the two-step melting region pinches off at a multicritical locus which separates the regime with two-step melting from a regime in which the melting is first order~\cite{cardy1980GeneralDiscrete,alcaraz1980,alcaraz1987}. A self-dual line intersects this multicritical locus, making it straightforward to tune to multicriticality by working with an explicitly self-dual lattice model~\cite{cardy1980GeneralDiscrete,alcaraz1980,alcaraz1987}. This multicriticality was argued~\cite{alcaraz1987,dorey1999PhaseDiagram} to be controlled by the $Z_6$ parafermionic CFT~\cite{zamolodchikov1985NonlocalParafermion} alluded to earlier.
On symmetry grounds, we may identify $\psi^m(r)$ in the Ising antiferromagnet with  $\exp(i m \theta(r))$ ($m=1,2,3$) in the lattice-discretized Landau theory.

\begin{figure}
    \centering
    \includegraphics[width=\columnwidth]{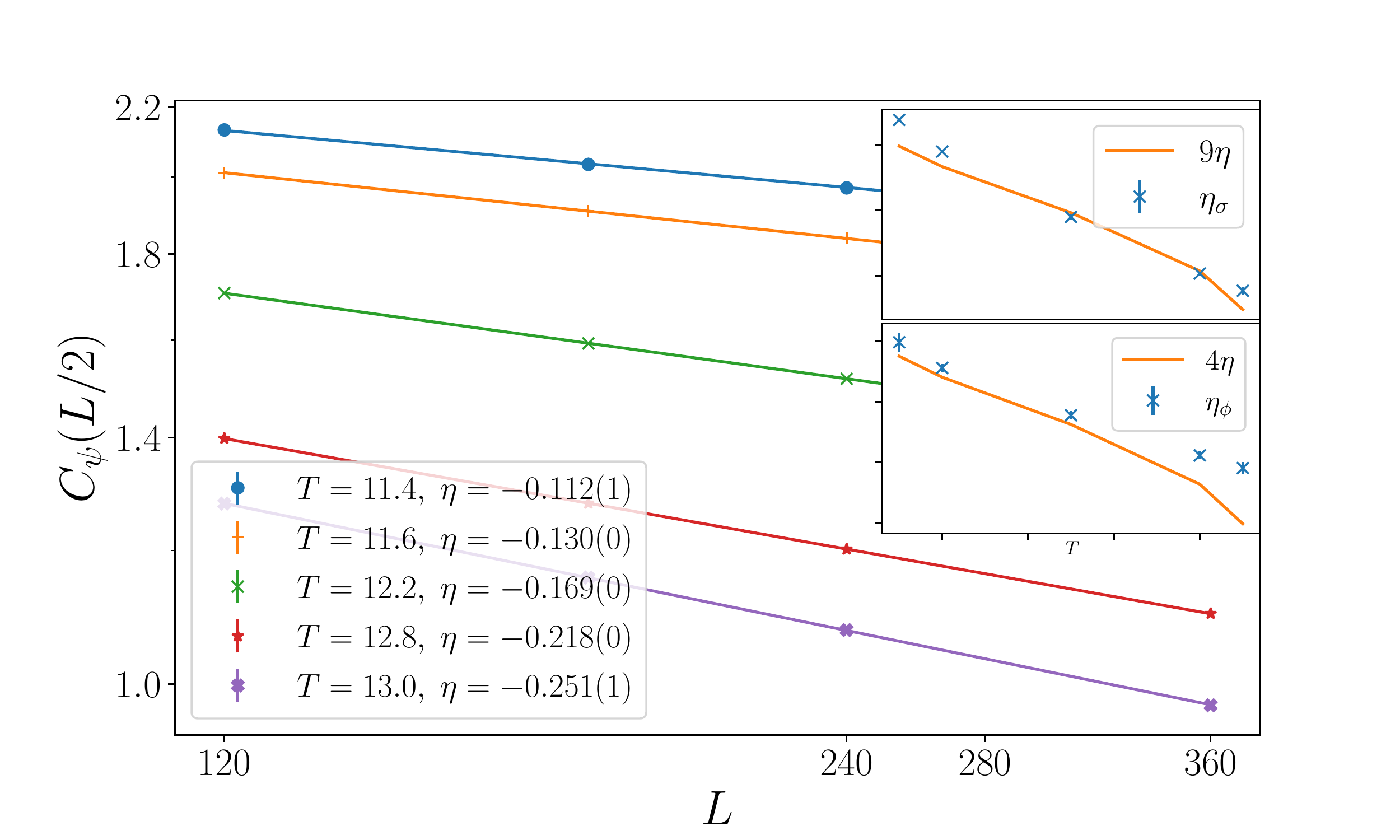}
    \caption{ The two-point correlation function of the three-sublattice order parameter $C_{\psi}(\vec{r})$ at separation $\vec{r} = \hat{e}_x \frac{L}{2}$ as a function of the system size $L$, for various values of temperatures in the intermediate power-law region ($T \in [5.7,6.5]$) at $R=2$ and $\kappa=3.5$. Power-law fits to $1/L^{\eta_{\psi}}$ are shown. Subplots show the critical exponents $\eta_{\phi}$ and $\eta_{\sigma}$ extracted from power-law fits to $C_{\phi}(L/2)$ and $C_{\sigma}(L/2)$ respectively in the same temperature range. $\eta_{\phi}$ and $\eta_{\sigma}$ are compared to the extracted values of $\eta_{\psi}$ via the relation $\eta_{\phi} = 4 \eta_{\psi}$ and $\eta_{\sigma} = 9 \eta_{\psi}$.}
    \label{fig:two_step}
\end{figure}

\begin{figure}
    \centering
    \includegraphics[width=\columnwidth]{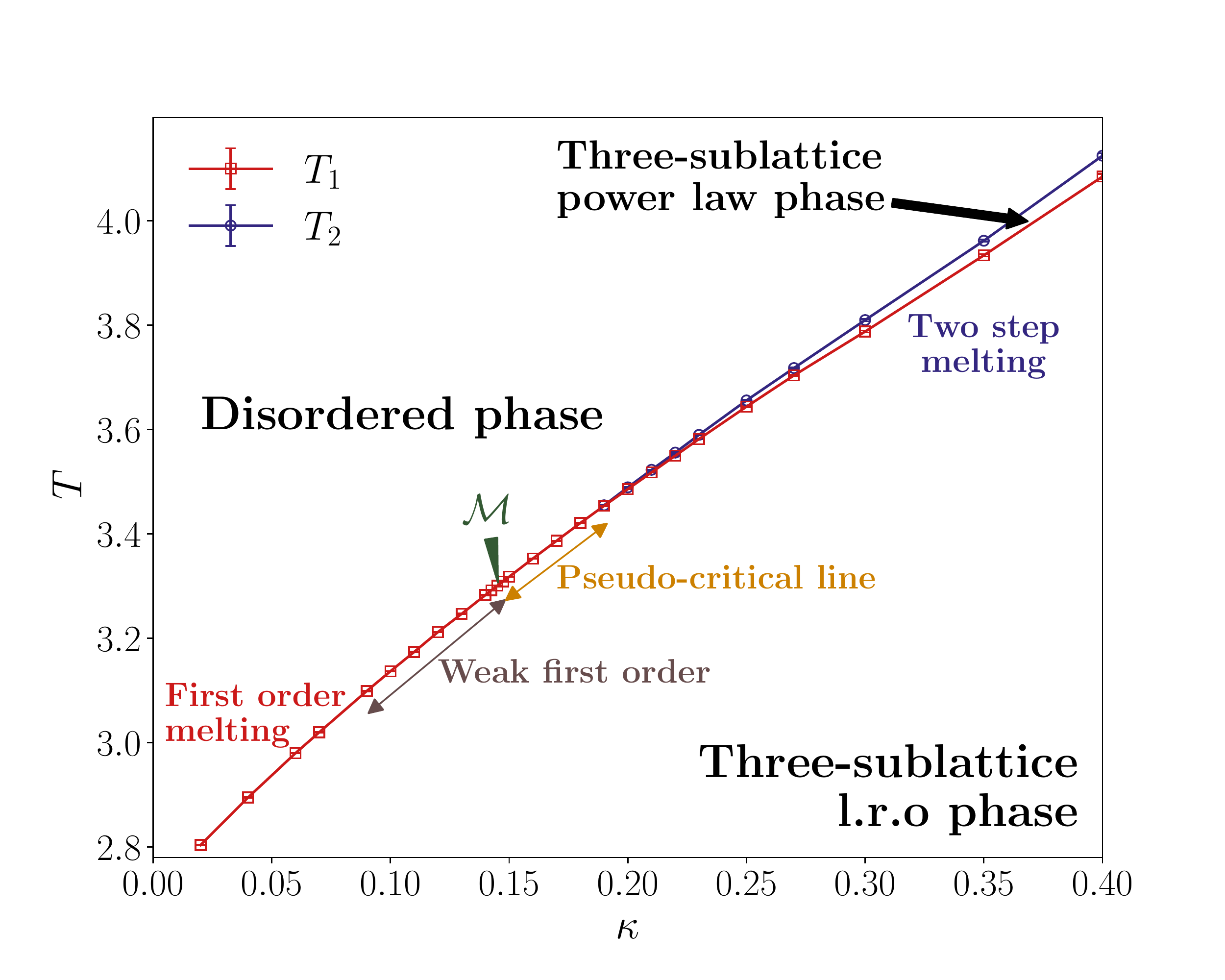}
    \caption{$\kappa-T$ phase diagram of the Ising model on the triangular lattice with antiferromagnetic nearest neighbor ($J_1$), ferromagnetic next-nearest neighbor $(J_2)$ and ferromagnetic next-next-nearest neighbor ($J_3$) interactions on the $R=2$ line. Cubic splines provide a guide to the eye. The multicritical point is at $\kappa \approx 0.1425$. It separates a pseudo-critical line from a  line of weakly first order transitions.}
    \label{fig:transition}
\end{figure}

{\em Two-step melting region:} Thus, in the case of the triangular lattice antiferromagnet in the intermediate power-law region we expect $C_{\psi}(\vec{r}) \sim 1/{r^{\eta_{\psi}}}$ with $\eta_{\psi} \equiv \eta_{\theta} \in [1/9,1/4]$. Similarly, we expect power-law behavior for the correlators of $\phi$ and $\sigma$, with the corresponding anomalous dimensions given by $\eta_{\phi} \equiv \eta_{2\theta} = 4\eta_{\theta}$ and $\eta_{\sigma} \equiv \eta_{3\theta} = 9\eta_{\theta}$. We have confirmed that this is indeed the case. An example of this behavior is displayed in Fig.{\ref{fig:two_step}} at $R=2$ and $\kappa=3.5$. Thus, the two-step melting region can be detected by measuring the critical exponents $\eta_{\psi}, \eta_{\phi}$ and $\eta_{\sigma}$. We use this as a guide to the width in $T$ of the power-law ordered intermediate phase. Fig.~\ref{fig:transition} shows the $\kappa-T$ melting phase diagram along the $R=2$ line.

{\em Pseudocritical behaviour and weakly first order line:} For $\kappa<0.2$ along this line, we cannot distinguish the two-step melting region any more. Indeed, the melting appears to proceed via a single continuous transition (Fig.~\ref{fig:transition}). This behavior, which we dub ``pseudocritical melting'' is ascribed to a two-step melting region that is too narrow in $T$ to resolve in our numerics. 
This interpretation of the pseudocritical regime, in terms of a two-step melting region that is too narrow to resolve, is reasonable since Landau theory arguments rule out a single continuous transitions except at a fine-tuned multicritical point. Interestingly, this pseudocritical behavior extends quite some distance away from this putative multicritical point. As we reduce $\kappa$ further, we find that the melting becomes weakly first order in nature. This is visible for $\kappa \lesssim 0.1425$ along the $R=2$ line.
Taken together, all of this points to the presence of a multicritical point that separates this pseudocritical behavior from weakly-first order behavior.  Similar behavior is found along the $Y=1.05$ line between the regime with two-step melting and the regime with weakly first order behavior. 

\begin{figure*}
    \centering
    \includegraphics[width=\textwidth]{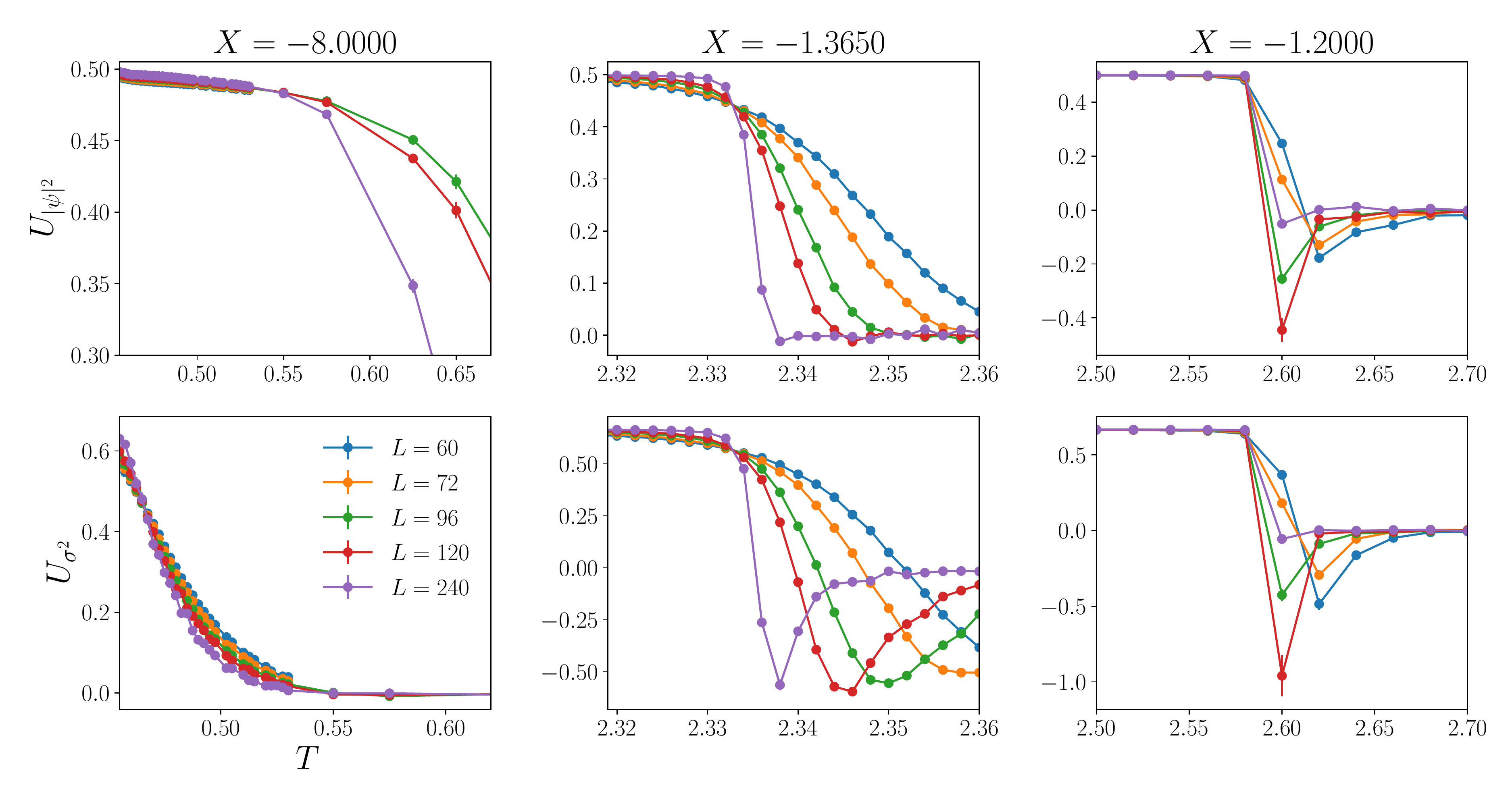}
    \caption{Binder cumulants $U_{\psi}$ and $U_{\sigma}$ on the $Y=1.05$ line at (left)$X=-8$, in the two-step melting region, at (middle)$X=-1.365$ at the approximate location of the multicritical point, and (right) at $X=-1.2$, where the transition is weakly first order.}
    \label{fig:binder_all}
\end{figure*}
{\em Binder cumulants:} To put these conclusions on a firmer footing, we track the behavior of the Binder cumulants $U_{\psi}$ and $U_{\sigma}$ along the $R=2$ line in the vicinity of the putative multicritical point, as well as along the $Y=1.05$ line in the vicinity of a similar multicritical point on this line as well as in the two-step melting region and the region with first-order behavior.
Fig.~\ref{fig:binder_all} shows the Binder cumulants $U_{\psi}$ and $U_{\sigma}$ for various values of $L$ at $X=-8,-1.365$ and $-1.2$ on the $Y=1.05$ line. At $X=-8$, the three-sublattice order undergoes a two-step melting. This is signalled by a region in which the Binder cumulant curves of different sizes stick together. At $X=-1.2$, when the melting is weakly first order, we find characteristic dips below zero in the Binder cumulants, with the size of the dip scaling with $L$. As one moves towards the two-step melting region, the dips in $U_{\psi}$ disappear first, leading to an extended region in which $U_{\sigma}$ continues to show dips below zero. However, these do not scale with $L$. Moving further towards the two-step melting region, these dips eventually disappear, and we start seeing the stick and splay behavior characteristic of the Binder cumulants in the vicinity of two-step melting. 
Clearly, this curious behavior of the Binder cumulants has its origins in the fact that the power-law ordered intermediate phase associated with two-step melting pinches off and is replaced by a weakly first order transition line. 

{\em Locating the multicritical point:} To obtain a quantitative estimate for the location of the associated multicritical point separating these two regimes, we need further input.
As we describe in more detail below, this additional input is provided by the entirely analogous behavior seen in the corresponding Binder cumulants along a self-dual locus that passes through a multicritical point in the phase diagram of the generalized six-state clock model. Along this self-dual locus, we find that the multicritical point of the generalized six-state clock model is associated with the threshold at which the dip in the Binder cumulant of $\exp(i \theta)$ disappears, leaving behind a dip in the Binder cumulant of $\exp(3 i\theta)$ which does not scale with $L$.
Since $\exp(i \theta)$ represents the three-sublattice order parameter $\psi$ within this Landau theory, we may locate the multicritical transition of the Ising antiferromagnet by identifying the threshold at which the dip in the Binder cumulant of of $\psi$ disappears. This gives the estimate $X \approx -1.365$ ($\kappa \approx 0.1425$) on the $Y= 1.05$ line (on the $R=2$ line).

\begin{figure*}
    \centering
    \includegraphics[width=\textwidth]{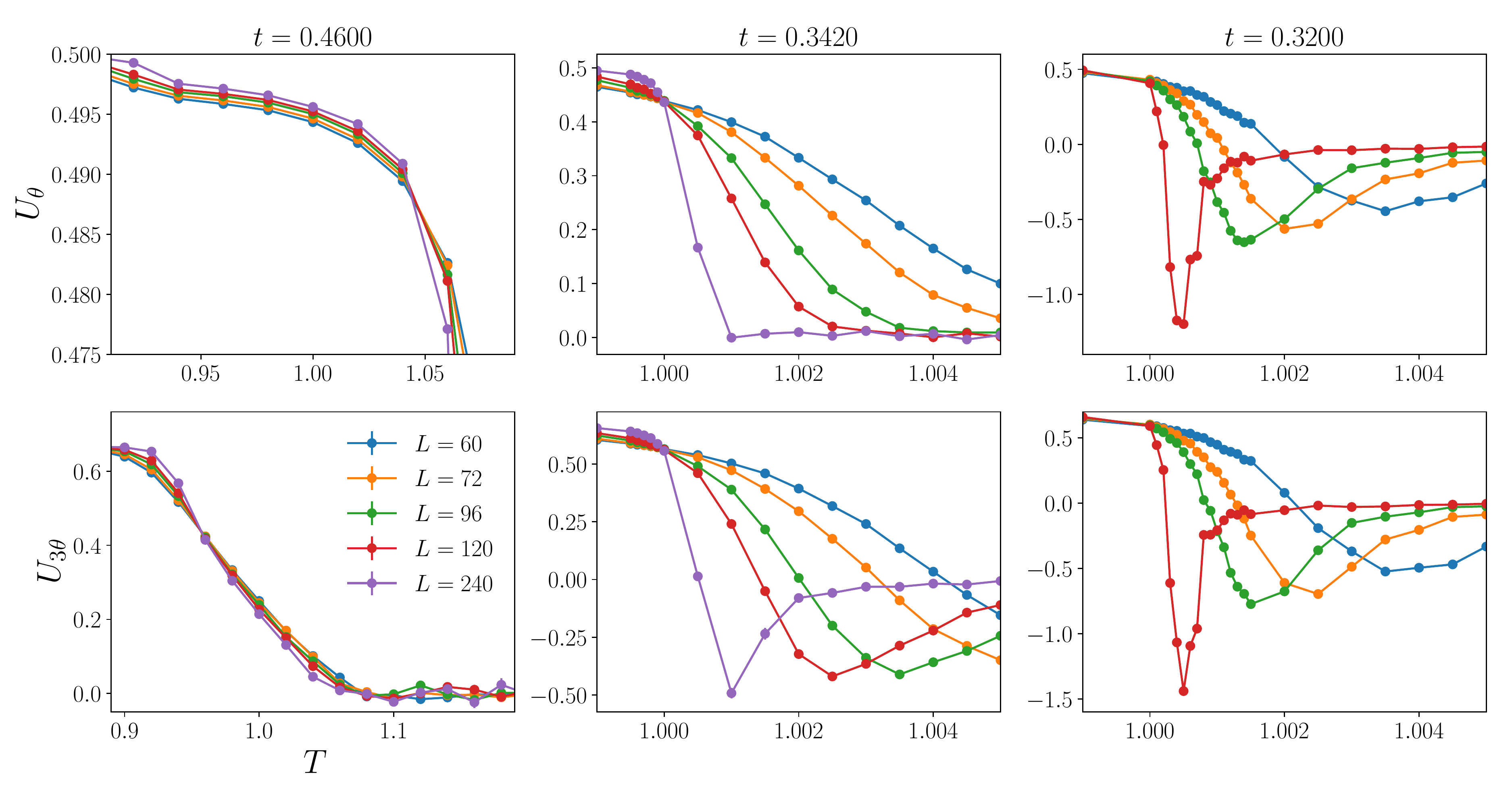}
    \caption{Binder cumulants $U_{\theta}$ and $U_{3\theta}$ (left) at $t=0.46$, in the two-step melting region, (middle) at the multicritical point $t_m \approx 0.342$   and (right) at $t=0.32$ when the transition is weakly first order.}
    \label{fig:binder_clock}
\end{figure*}

{\em Inputs from $Z_6$ clock model along the self-dual line:}
As already mentioned above, this identification of multicritical melting is based on the behavior along the self-dual locus in the phase diagram of the generalized six-state clock model. We follow Ref.~\cite{dorey1999PhaseDiagram} and set the interactions to values $K_1(t)$, $K_2(t)$, and $K_3(t)$~\cite{cardy1980GeneralDiscrete,alcaraz1980,alcaraz1987,dorey1999PhaseDiagram} parameterized by a dimensionless coordinate $t$ that represents the position along the self-dual locus $(t,T=1)$~\cite{dorey1999PhaseDiagram}. Since this locus is known to pass through a multicritical point that separates two-step melting behavior from a first order melting transition~\cite{cardy1980GeneralDiscrete,alcaraz1980,alcaraz1987,dorey1999PhaseDiagram}, this choice of couplings guarantees that the multicritical point will be at $T=1$, $t=t_m$ in these units. To identify $t_m$, we just need to monitor the behavior in the vicinity of $T=1$, and identify $t_m$ with the point at which long range order is lost via a single continuous transition with transition temperature $T_c=1$ in these units. For $t > t_m$, the system melts via a two-step melting process, with $T=1$ lying within the power-law ordered phase, while for $t < t_m$ the transition is first order in nature~\cite{cardy1980GeneralDiscrete,alcaraz1980,alcaraz1987,dorey1999PhaseDiagram}. 

Since we know that the multicritical point must be at $T_c=1$, a study of the Binder cumulants near the melting transitions along this self-dual line labeled by $t$ gives us information on the expected behavior of Binder cumulants in the vicinity of the multicritical point.  Identifying $t_m \approx 0.342$ to be the value of $t$ at which the clock model melts via a single continuous transition at $T_c = 1$, we see (Fig.~\ref{fig:binder_clock}) that the Binder cumulants $U_{\theta}$ and $U_{3\theta}$ have behavior entirely analogous to that encountered earlier near multicriticality in the Ising antiferromagnet: At $t=0.46$ in the two-step melting regime, at and near $t=t_m$, and at $t=0.32$ in the first order melting region, we see that the Binder cumulants behave exactly as they do in the corresponding regimes of the phase diagram of the triangular Ising antiferromagnet. Indeed, we see that it becomes difficult to resolve the two-step melting region as one approaches $t_m$; instead, one sees pseudocritical behavior. This is essentially identical to the behavior we have studied in the triangular Ising antiferromagnet. 

\begin{figure}
    \centering
    \includegraphics[width=\columnwidth]{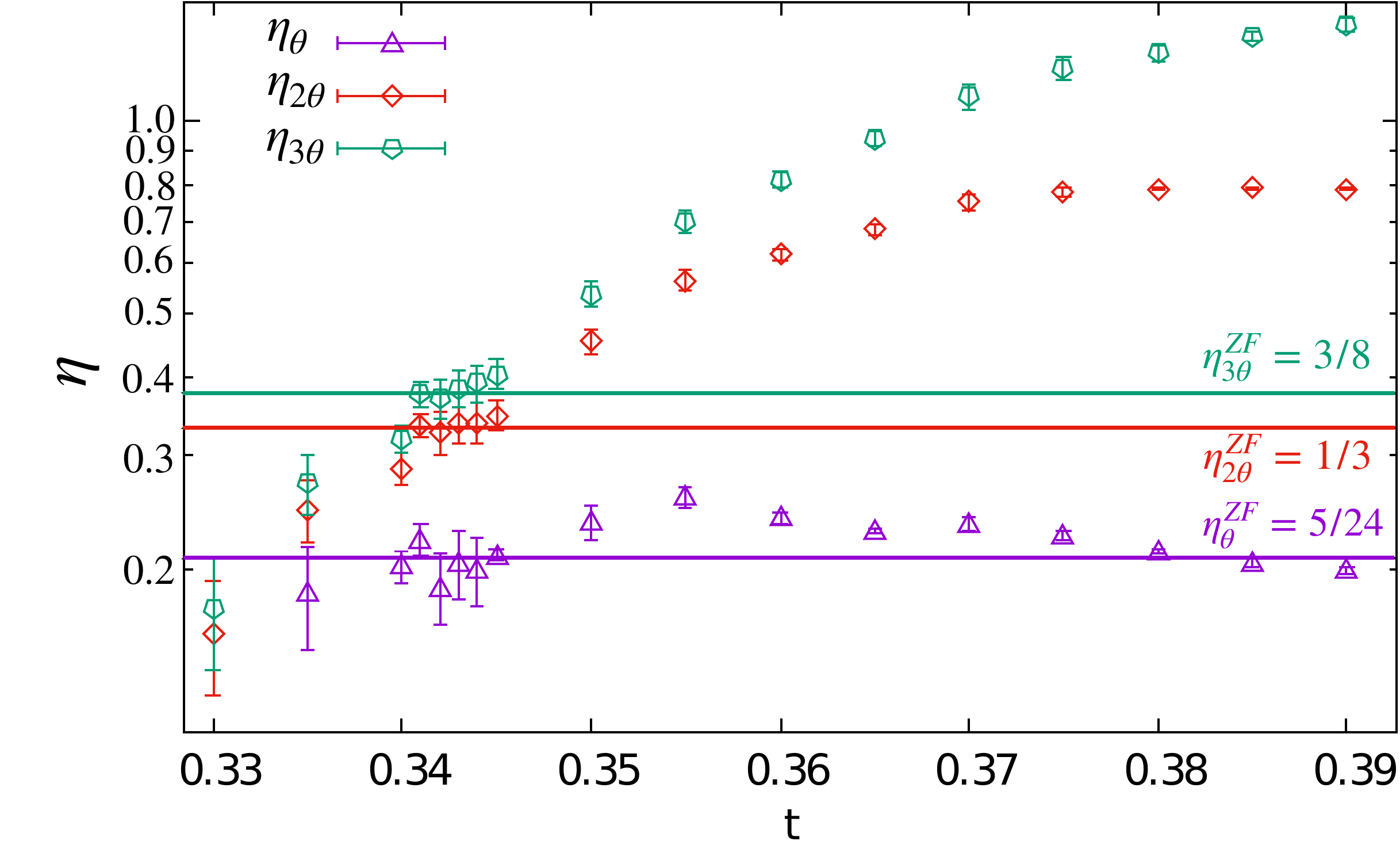}
    \caption{ Values of $\eta_{\theta}$, $\eta_{2 \theta}$ and $\eta_{3 \theta}$ measured as a function of $t$ in the six-state clock model. The solid lines represent the value of $\eta$ as predicted by the $Z_6$ self-dual parafermionic conformal field theory constructed by Zamolodchikov and Fateev\cite{zamolodchikov1985NonlocalParafermion}. The exponents converge to the predicted values at the multicritical point $t \approx 0.342$.}
    \label{fig:eta_vs_t}
\end{figure}

\begin{figure}
    \centering
    \includegraphics[width=\columnwidth]{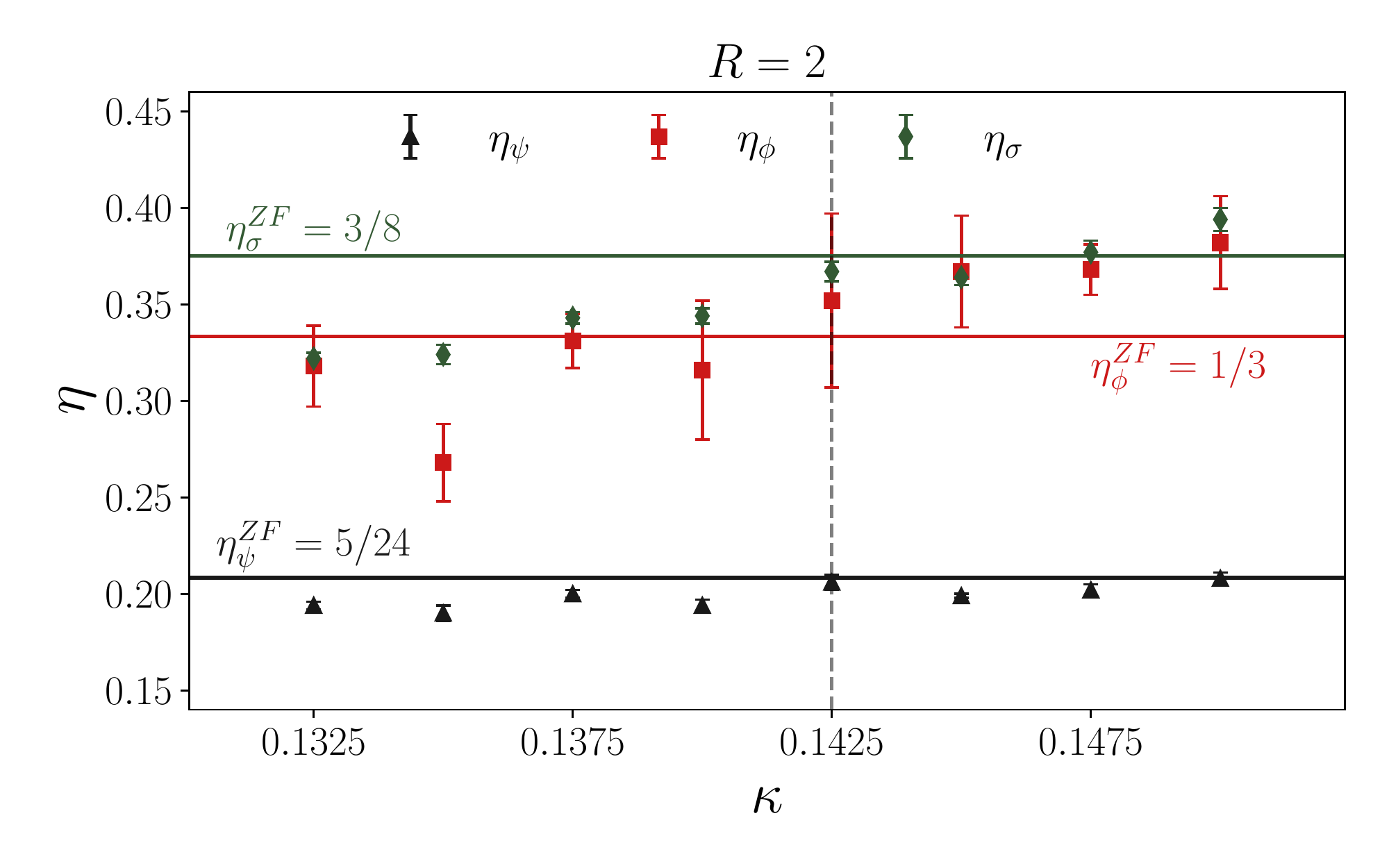}
    \caption{Values of $\eta_{\psi}$, $\eta_{\phi}$ and $\eta_{\sigma}$ measured on the $R=2$ line for $\kappa \in [0.1325,0.15]$ in the triangular lattice antiferromagnet. The solid lines represent the value of $\eta$ as predicted by the $Z_6$ self-dual parafermionic conformal field theory constructed by Zamolodchikov and Fateev\cite{zamolodchikov1985NonlocalParafermion}. The exponents converge to the predicted values at the multicritical point $\kappa \approx 0.1425$.}
    \label{fig:eta_vs_kappa}
\end{figure}

{\em Universality class of multicritical locus:} By way of numerical evidence for the predicted~\cite{alcaraz1987,dorey1999PhaseDiagram} universality class of the multicritical point at $t_m=0.342$, $T_c=1$, we plot the measured exponents $\eta_{\theta},\eta_{2\theta}$ and $\eta_{3\theta}$ at the pseudocritical transition point for various values of $t$ around $t=0.342$ in Fig.~\ref{fig:eta_vs_t}. It can be seen that the exponents approach the known values~\cite{zamolodchikov1985NonlocalParafermion} of 5/24,1/3 and 3/8 respectively in the $Z_6$ parafermionic universality class. 
Crucially, when we plot the exponents $\eta_{\psi}, \eta_{\phi}$ and $\eta_{\sigma}$ along the pseudocritical line for various values of $\kappa$ around the $\kappa=0.1425$ multicritical point on the $R=2$ line in the phase diagram of the triangular Ising antiferromagnet, we see that these exponents also approach these values characteristic of the self-dual $Z_6$ parafermionic CFT (Fig.~\ref{fig:eta_vs_kappa}). This provides further confirmation of our central thesis: The multicritical point separating two-step melting of three-sublattice order from first order melting behavior in the phase diagram of triangular Ising antiferromagnets provides a physical realization of the self-dual $Z_6$ parafermionic CFT~\cite{zamolodchikov1985NonlocalParafermion}.

{\em Discussion and outlook:} Several earlier proposals for potential realizations of supersymmetric CFTs in experimental systems have focused on the possibility of accessing such behaviors in the ground-state phase diagram of quantum systems~\cite{grover2014EmergentSpaceTime,li2017EdgeQuantum,rahmani2015EmergentSupersymmetry,dalmonte2015Cluster,alberton2017Fate,li2018NumericalObservation,bauer2013Supersymmetric,ma2021RealizationSupersymmetry}. When such behavior is associated with boundaries of higher dimensional topological phases of matter, it can sometimes be accessed by tuning just one parameter. Nevertheless, independent of the number of parameters that need to be tuned, such putative experimental realizations rely heavily on having controlled access to quantum coherent behaviour at ultra-low temperatures. In contrast, the $Z_6$ parafermion multicriticality identified here is associated with the melting of three-sublattice order in triangular lattice Ising antiferromagnets at relatively high temperatures that are of the same order as the exchange couplings. Although some degree of fine-tuning is needed to access the vicinity of this multicritical threshold, this multicriticality influences the melting behavior over a relatively wide range of parameters, leading to the pseudo-critical behavior described here. This can be detected in simple thermodynamic measurements of the susceptibility to a uniform external field along the easy axis, making this interesting multicritical behavior that much more accessible to experiment. It would therefore be interesting to identify candidate triangular lattice antiferromagnets with strong easy-axis anisotropy and further neighbour interactions that stabilize a three-sublattice ordered low temperature state, and understand their observed finite temperature behavior in terms of the phase diagram described here.

{\em Acknowledgments:} We acknowledge useful discussions with D. Dhar, D. Heidarian,  R. Loganayagam, P. Narayan, N. Shannon and N. Karthik, and computer cluster related assistance from K. Ghadially and A. Salve of the Department of Theoretical Physics (DTP) of the Tata Institute of Fundamental Research (TIFR). GR also gratefully acknowledges technical assistance from the Scientific Computing and Data Analysis section of the Okinawa Institute of Science and Technology (OIST). A significant part of the work presented here contributed to the Ph.D thesis submission of GR to the TIFR Deemed University (2019). ND's work formed a part of their undergraduate thesis (2014) submitted to the Birla Institute of Technology and Science Pilani , K.K Birla Campus, Goa (BITS-Goa). SS's work formed part of their undergraduate thesis submitted to the Indian Institute of Technology Bombay (IITB). ND and SS gratefully acknowledge the hospitality of DTP, TIFR during their undergraduate thesis work. This work was made possible by the generous allocation of computational resources made by DTP, TIFR, as well as by OIST. GR was supported by a graduate fellowship of the TIFR during a significant part of this work, and by the Theory of Quantum Matter unit at the OIST during the rest of this work. ND was supported at the TIFR by a National Postdoctoral Fellowship of SERB, DST India (NPDF/2020/001658) during the preparation of this manuscript. SS was supported by a graduate fellowship at Princeton University during the preparation of this manuscript. KD was supported at the
TIFR by DAE, India and in part by a J.C. Bose Fellowship
(JCB/2020/000047) of SERB, DST India, and by
the Infosys-Chandrasekharan Random Geometry Center
(TIFR).

{\em Author contributions:} GR performed all the computations with the assistance of ND (for the Ising models) and SS (for the $Z_6$ clock models). KD conceived and directed the computational work, and finalized the manuscript using detailed inputs from GR.

\bibliography{references}

\end{document}